\documentstyle[preprint,eqsecnum,aps]{revtex}
\begin{document}
\draft
\widetext
\title{Colliding waves in metric--affine gravity}
\author{
Alberto Garc\'{\i}a$^a$\thanks{E-mail: aagarcia@fis.cinvestav.mx},
Claus L\"ammerzahl$^b$\thanks{E-mail:
claus@spock.physik.uni-konstanz.de}, \\
Alfredo Mac\'{\i}as$^c$\thanks{E-mail: amac@xanum.uam.mx},
Eckehard W. Mielke$^c$\thanks{E-mail: ekke@xanum.uam.mx},
and Jos\'e Socorro$^d$\thanks{E-mail: socorro@ifug4.ugto.mx}\\
$^a$  Departamento de F\'{\i}sica,\\
CINVESTAV--IPN, Apartado Postal 14--740, C.P. 07000, M\'exico, D.F.,
MEXICO\\
$^b$ Fakult\"at f\"ur Physik, Universit\"at Konstanz \\ Postfach 5560
M674, D--78434 Konstanz, Germany \\
$^c$ Departamento de F\'{\i}sica,\\
Universidad Aut\'onoma Metropolitana--Iztapalapa,\\
Apartado Postal 55-534, C.P. 09340, M\'exico, D.F., MEXICO.\\
$^d$ Instituto de F\'{\i}sica de la Universidad de Guanajuato,\\
Apartado Postal E-143, C.P. 37150, Le\'on, Guanajuato, MEXICO.
}

\date{\today}

\maketitle

\begin{abstract}
We generalize the formulation of the colliding gravitational waves to
metric--affine
theories and present an example of such kind of exact solutions. The
plane
waves are equipped with five symmetries and the resulting
geometry after the collision possesses two spacelike Killing
vectors.
\end{abstract}
\vspace{0.5cm}

\pacs{PACS numbers: 04.50.+h; 04.20.Jb; 03.50.Kk}

\narrowtext

%**********************************************************
\section{Introduction}

In recent years, the collision of plane--fronted gravitational waves
possibly coupled with electromagnetic waves, has been extensively
studied \cite{khan,szek,chandra1,chandra2}. Because gravity is always
attractive, it was expected that focusing of the waves would occur and one of 
the interesting questions is how much focusing does general relativity predict.
Within this framework, strong focusing would appear by the development of
spacetime curvature singularities.
Many solutions has been presented so far, describing the collisions of 
plane--fronted gravitational and electromagnetic waves. And quite a few of 
them do develop Cauchy horizons.

The spacetimes describing the interaction region produced after the
collision of plane gravitational waves contain two spacelike
Killing vectors and there exist several generating techniques to
obtain solutions with these symmetries. All the techniques developed for
stationary axysimmetric spacetimes can be applied to generate cylindrically
symmetric spacetimes, in particular, colliding plane waves.

On the other hand, if one gauges the affine group and additionally 
allows for a metric $g$, then one ends up with the metric--affine 
gauge theory of gravity (MAG)\cite{PR}.
The four--dimensional affine group $A(4,{\bf R})$ is the
semidirect product of the {\it translation} group $R^4$ and the 
{\em linear group} 
$GL(4,{\bf R})=R^+ \otimes [T\, {\times\!\!\!\!\!\!\!\subset}\, 
SL(4\,,{\bf R})]$. 
This spacetime encompasses two different
post--Riemannian structures: the nonmetricity one--form
$Q_{\alpha\beta}=Q_{i\alpha\beta} \,dx^i$ and the torsion two--form
$T^\alpha =\frac{1}{2}\,T_{ij}{}^\alpha dx^i\wedge dx^j$.
In the Yang-Mills fashion, gauge Lagrangians quadratic in
curvature, torsion, and nonmetricity are considered.
One way to investigate the potentialities of such models is to
look for {\em exact} solutions.

The search for exact solutions within MAG has been pioneered by
Tres\-guerres\cite{Tres14,Tres15} and by Tucker and Wang\cite{TW}.
With propagating nonmetricity $Q_{\alpha\beta}$, two types of charge
are expected to arise: {\em One dilation charge} related by Noether
procedure to the trace $Q:=Q_\gamma{}^\gamma/4$ of the nonmetricity, called the
Weyl covector $Q=Q_i dx^i$. It is the connection associated with gauging
${\bf R}^+$ instead of $U(1)$ for the Maxwell potential $A=A_i dx^i$.
{\em Nine} types of {\em shear charge} related to the remaining
traceless piece 
${\nearrow\!\!\!\!\!\!\!Q}_{\alpha\beta}:=Q_{\alpha\beta}-Q\,g_{\alpha\beta}$
of the nonmetricity.  Under the local Lorentz group, the nonmetricity can be 
decomposed into four irreducible pieces $^{(I)}Q_{\alpha\beta}$, with 
$I=1,2,3,4$. The Weyl covector is linked to 
$^{(4)}Q_{\alpha\beta}=Q\,g_{\alpha\beta}$.

The following natural step in these lines is to elucidate the
behavior of interacting plane waves. Of particular relevance is the head--on
collision of two plane waves, i.e. the colliding wave problem.
It is assumed that in the corresponding spacetime, the two waves
approach each other, from opposite sides, in flat Minkowski background; after
the collision, a new gravitational field evolves, which satisfies certain
continuity conditions. The plane waves are equipped with five
symmetries, while the geometry resulting after the collision possesses two
spacelike Killing vectors.
The main pruporse of this paper is to generalize the formulation of
the colliding waves to MAG theories and to present an example of such
kind of solutions. We will take advantage of the fact that
certain MAG models can be reduced to an {\em effective}
Einstein--Proca system\cite{De96}. Mac\'{\i}as et al. \cite{ma}, and Socorro 
et al. \cite{slmm} mapped the Einstein--Maxwell sector of the dilaton--gravity
coming from low energy string theory, to MAG, thus, finding soliton and 
multipole solutions. 

The plan of the paper is as follows: In Sec. 2 the quadratic MAG
Lagrangian is revisited. In Sec. 3 the generalization of the colliding waves
concept to MAG is developed.  In Sec. 4 a colliding wave solution in MAG is
presented. In Sec. 5 the results are discussed.

%*******************************************
\section{Quadratic MAG Lagrangian}

In a metric--affine spacetime, the curvature has {\em eleven} irreducible
pieces, see \cite{PR}, Table~4.  If in addition we recall that the nonmetricity
has {\em four} and the torsion {\em three} irreducible pieces, then a
general quadratic Lagrangian in MAG reads:
\begin{eqnarray}
\label{QMA} V_{\rm MAG}&=&
\frac{1}{2\kappa}\,\left[-a_0\,R^{\alpha\beta}\wedge
\eta_{\alpha\beta}
-2\lambda\,\eta+
T^\alpha\wedge{}^*\!\left(\sum_{I=1}^{3}a_{I}\,^{(I)}
T_\alpha\right)\right.\nonumber\\
&+&\left.  2\left(\sum_{I=2}^{4}c_{I}\,^{(I)}Q_{\alpha\beta}\right)
\wedge\vartheta^\alpha\wedge{}^*\!\, T^\beta + Q_{\alpha\beta}
\wedge{}^*\!\left(\sum_{I=1}^{4}b_{I}\,^{(I)}Q^{\alpha\beta}\right)\right]
\nonumber\\&- &\frac{1}{2}\,R^{\alpha\beta} \wedge{}^*\!
\left(\sum_{I=1}^{6}w_{I}\,^{(I)}W_{\alpha\beta} +
  \sum_{I=1}^{5}{z}_{I}\,^{(I)}Z_{\alpha\beta}\right)
\label{lobo}\,.
\end{eqnarray}
In the above, the Minkowsi metric is $o_{\alpha\beta} = 
\hbox{diag}(-+++)$, $\eta:={}^*\!\, 1$ is the
volume four--form and the
constants $a_0,\cdots a_3$, $b_1,\cdots b_4$, $c_2, c_3,c_4$,
$w_1,\cdots w_6$, $z_1,\cdots z_5$ are dimensionless. In the curvature square
term we have introduced the irreducible pieces of the antisymmetric part
$W_{\alpha\beta}:= R_{[\alpha\beta]}$ and the symmetric part
$Z_{\alpha\beta}:= R_{(\alpha\beta)}$ of the curvature two--form.
Again, in $Z_{\alpha\beta}$, we meet a purely post--Riemannian part.
The segmental curvature $^{(4)}Z_{\alpha\beta}:=
R_\gamma{}^\gamma\,g_{\alpha\beta}/4= g_{\alpha \beta} dQ$ has
formally a similar structure as the electromagnetic field strength $F=dA$.

Let us recall the three general field equations of MAG,
see \cite{PR} Eqs.(5.5.3)--(5.5.5). Because of its redundancy, we
omit the zeroth field equation with its gauge momentum $M^{\alpha\beta}$.
The first and the second field equations read
\begin{eqnarray}
DH_{\alpha}-
E_{\alpha}&=&\Sigma_{\alpha}\,,\label{first}\\
DH^{\alpha}{}_{\beta}-
E^{\alpha}{}_{\beta}&=&\Delta^{\alpha}{}_{\beta}\,,
\label{second}
\end{eqnarray}
where $\Sigma_{\alpha}$ and $\Delta^{\alpha}{}_{\beta}$ are the
canonical energy--momentum and hypermomentum current three--forms
associated with matter. We will consider the {\em vacuum case} with
$\Sigma_{\alpha}=\Delta^{\alpha}{}_{\beta}=0$. The left hand sides
of (\ref{first})--(\ref{second}) involve the gravitational gauge field
momenta two-forms $H_{\alpha}$ and $H^{\alpha}{}_{\beta}$
(gravitational ``excitations"). We find them, together with
$M^{\alpha\beta}$, by partial differentiation of the Lagrangian (\ref{QMA}):
\begin{eqnarray}
M^{\alpha\beta}&:=&-2{\partial V_{\rm MAG}\over \partial
Q_{\alpha\beta}}=
-{2\over\kappa}\Bigg[{}^*\! \left(\sum_{I=1}^{4}b_{I}{}^{(I)}
Q^{\alpha\beta}\right)\nonumber\\
&& + c_{2}\,\vartheta^{(\alpha}\wedge{}^*\! ^{(1)}T^{\beta)} +
c_{3}\,\vartheta^{(\alpha}\wedge{}^*\! ^{(2)}T^{\beta)} +
{1\over 4}(c_{3}-c_{4})\,g^{\alpha\beta}{}^*\!\,
T\Bigg]\,,\label{M1}\\
  H_{\alpha}&:=&-{\partial V_{\rm MAG}\over \partial T^{\alpha}} = -
  {1\over\kappa}\,
  {}^*\!\left[\left(\sum_{I=1}^{3}a_{I}{}^{(I)}T_{\alpha}\right) +
    \left(\sum_{I=2}^{4}c_{I}{}^{(I)}
Q_{\alpha\beta}\wedge\vartheta^{\beta}\right)\right],\label{Ha1}\\
  H^{\alpha}{}_{\beta}&:=& - {\partial V_{\rm MAG}\over \partial
    R_{\alpha}{}^{\beta}}= {a_0\over
2\kappa}\,\eta^{\alpha}{}_{\beta} +
  {\cal W}^{\alpha}{}_{\beta} + {\cal Z}^{\alpha}{}_{\beta},\label{Hab1}
\end{eqnarray}
where we introduced the abbreviations
\begin{equation}
  {\cal W}_{\alpha\beta}:= {}^*\!
  \left(\sum_{I=1}^{6}w_{I}{}^{(I)}W_{\alpha\beta}
\right),\quad\quad
  {\cal Z}_{\alpha\beta}:= {}^*\!
  \left(\sum_{I=1}^{5}z_{I}{}^{(I)}Z_{\alpha\beta} \right).
\end{equation}

Finally, the three--forms $E_{\alpha}$ and $E^{\alpha}{}_{\beta}$
describe the canonical energy--mo\-men\-tum and hypermomentum
currents of the gauge fields themselves. One can write them as follows
\cite{PR}:
\begin{eqnarray}
E_{\alpha}&=&e_{\alpha}\rfloor V_{\rm MAG} + (e_{\alpha}\rfloor
T^{\beta})
\wedge H_{\beta} + (e_{\alpha}\rfloor R_{\beta}{}^{\gamma})\wedge
H^{\beta}{}_{\gamma} + {1\over 2}(e_{\alpha}\rfloor Q_{\beta\gamma})
M^{\beta\gamma}\, , \\
E^{\alpha}{}_{\beta}&=& - \vartheta^{\alpha}\wedge H_{\beta} -
M^{\alpha}{}_{\beta}\, ,
\end{eqnarray}
where $e_{\alpha}\rfloor$ denotes the interior product with the
frame.

%********************************************************
\section{Colliding waves in MAG}

This work, as was stated previously, is concerned with fields
interpretable as colliding wave solution. With this goal in mind, we extend the
definition of vacuum colliding waves, defined by Ernst et al. \cite{ernst} to 
MAG theories.

The set of colliding waves solutions in metric--affine gravity
is described by the metric
\begin{equation}
g = 2 g(u,v)\, du \, dv + g_{ab}(u,v) dx^a dx^b \, , \qquad\,
a,b=1,2\, ,
\end{equation}
which only depends on the advanced and retarded time $u := t - z$ and 
$v := t + z$, respectively.
The domain of the coordinate charts consists of $(x,y) \in {\bf R}^2$
and $(u,v)\in {\bf R}^2$; it is the union of four continuous
regions: $\hbox{I} := \{(u,v): 0\leq u < 1, 0\leq v < 1\}$ ,
$\hbox{II} := \{(u,v): u < 0,
0\leq v < 1\}$, $\hbox{III} := \{(u,v): 0\leq u < 1, v < 0\}$,
$\hbox{IV} :=\{(u,v): u\leq 0, v\leq 0\}$, see Fig.1.

As for the torsion and nonmetricity field configurations, we
concentrate on the simplest non--trivial case with shear. According to its
irreducible decomposition (see the Appendix B of \cite{PR}), the
nonmetricity contains two covector pieces, namely $^{(4)}Q_{\alpha\beta}=
Q\,g_{\alpha\beta}$, the dilation piece, and
\begin{equation}
  ^{(3)}Q_{\alpha\beta}={4\over
    9}\left(\vartheta_{(\alpha}e_{\beta)}\rfloor \Lambda - {1\over
      4}g_{\alpha\beta}\Lambda\right)\,,\qquad \hbox{with}\qquad
  \Lambda:= \vartheta^{\alpha}e^{\beta}\rfloor\!
  {\nearrow\!\!\!\!\!\!\!Q}_{\alpha\beta}\label{3q}\,,
\end{equation}
a proper shear piece. Accordingly, our ansatz for the nonmetricity
reads
\begin{equation}
  Q_{\alpha\beta}=\, ^{(3)}Q_{\alpha\beta} +\,
  ^{(4)}Q_{\alpha\beta}\,.\label{QQ}
\end{equation}
The torsion, in addition to its tensor piece,
encompasses a covector and an axial covector piece. Let us choose
only the covector piece as non-vanishing:
\begin{equation}
T^{\alpha}={}^{(2)}T^{\alpha}={1\over 3}\,\vartheta^{\alpha}\wedge
T\,,
\qquad \hbox{with}\qquad T:=e_{\alpha}\rfloor
T^{\alpha}\,.\label{TT}
\end{equation}
Thus we are left with the three non--trivial one--forms $Q$,
$\Lambda$, and $T$. We shall assume that this triplet of one--forms
share the spacetime symmetries, i.e. they depend on the variables $u$ and $v$
only.
The metric and the triplet fields have to be continuous over the
whole domain.

In the region IV, a subregion of the Minkowski space, it is
required that
\begin{equation}
g_{\mu\nu}(u,v)= g_{\mu\nu}(0,0)\, , \quad\,  Q= Q_0\, , \quad \,
\Lambda=\Lambda_0\, , \quad \, T=T_0\, ,
\end{equation}
which by scale transformations can be brought to standard
Minkowski metric and vanishing constants.
In region II, the metric components  and the triplet of one--forms
depends only on $v$, i.e. $g_{\mu\nu}=g_{\mu\nu}(0,v), Q=(0,v),
\Lambda=\Lambda(0,v)$, and $T=T(0,v)$. In region III these fields
are functions of the coordinate $u$, i.e. $g_{\mu\nu}=g_{\mu\nu}(u,0),
Q=(u,0), \Lambda=\Lambda(u,0)$, and $T=T(u,0)$. In region I, which is
occupied by the scattered null fields, the metric components and the triplet 
are functions of both $u$ and $v$ coordinates.

The metric, the torsion and the nonmetricity fields in region II and III
depend only on one variable, i.e. $u$ and $v$, respectively. Each of these 
regions
is equipped with five Killing vectors related with the metric. Moreover, the 
conformal Weyl tensor part corresponding to the Riemannian part possesses a 
quadrupole null eigendirection being covariantly constant. These two 
properties are characteristic of ppN waves. In region II, we have a pp wave, 
depending only on $v$, propagating to the right, while in region III the pp 
wave, depending on $u$, propagates to the left. Both waves collide at the 
event $u=v=0$, and from this event arises the interaction region I. 
In our case the torsion and nonmetricity depend in the various regions 
considered in the same way on $u$ and $v$ as the metric. 
Therefore, our situation describe also torsional and nonmetricity waves 
which propagate along null directions in regions II and III and collide 
in region I.

The following ansatz turns to be compatible with the above considerations,
\begin{equation}
Q=k_0\,
\rho(u,v)\,\vartheta^{\hat{2}}=\frac{k_0}{k_1}\Lambda
=\frac{k_0}{k_2}T
\label{genEug}\, .
\end{equation}
Here we introduced a second function $\rho(u,v)$ which has to be
determined by the field equations of MAG.

If we take the trace of the zeroth Bianchi identity
\begin{equation}
DQ_{\alpha\beta} = 2 Z_{\alpha\beta}
\label{0Bianchi}\, ,
\end{equation}
it merely consists of one irreducible piece $2dQ = Z_\gamma{}^\gamma
= \,^{(4)}Z_\gamma{}^\gamma$.
Consequently, $Q$ serves as a {\em potential} for
$^{(4)} Z_\gamma{}^\gamma$ in the same way as $A$ for $F=dA$.
In addition, the third part of (\ref{0Bianchi}) reads
$^{(3)}(DQ_{\alpha\beta}) = 2\,^{(3)}Z_{\alpha\beta}$, where
\begin{equation}
  ^{(3)}Z_{\alpha\beta}={2\over 3}\left(\vartheta_{(\alpha}\wedge
    e_{\beta)}\rfloor\delta - {1\over
      2}g_{\alpha\beta}\delta\right)\,,
  \quad\hbox{with}\quad\delta:={1\over 2}\vartheta^{\alpha}\wedge
e^{\beta}\rfloor\!{\nearrow\!\!\!\!\!\!\!Z}_{\alpha\beta}\,.\label{3z}
\end{equation}
The similarity in structure of (\ref{3q}) and (\ref{3z}) is
apparent.
Indeed, provided torsion carries only a covector piece,
see (\ref{TT}), we find
\begin{equation}
\delta={1\over 6}\,d\Lambda\,,\label{ddl}
\end{equation}
i.e.\ $^{(3)}Q_{\alpha\beta}$ acts as a potential for
$^{(3)}Z_{\alpha\beta}$.

In this way, the problem is reduced to
know the metric (coframe) and the fuction $\rho$.
Thus, the most general form of our fields compatible with colliding
wave spacetime structure is given by\cite{alberto1,alberto}:
\begin{eqnarray}
\rho&=&\rho(u,v)  , \, \,
C^\star_{abcd} = 2\Psi_0 U_{ab} U_{cd} + 2\Psi_2 (U_{ab} V_{cd}
+ V_{ab} U_{cd}+ W_{ab} W_{cd})+2\Psi_4 V_{ab} V_{cd}
\, , \quad {\rm region\, I} , \nonumber \\
\rho &=&\rho(v) \, , \quad C^\star_{abcd} = 2\Psi_0 U_{ab} U_{cd}\,
,\quad
{\rm region\, II}\, ,\nonumber\\
\rho &=&\rho(u)\, , \quad C^\star_{abcd} = 2 \Psi_4 V_{ab}  V_{cd}\,
,\quad
{\rm region\, III} \, ,
\end{eqnarray}
where $C^\star_{abcd}$ is the conformal Weyl tensor corresponding to the 
Riemannian part of the curvature tensor, and with
\begin{eqnarray}
W_{ab}&=&  m_a \tilde m_b - m_b \tilde m_a - k_a l_b+ k_bl_a \,
,\nonumber\\
V_{ab}&=&  k_a  m_b - k_b  m_a \,  , \nonumber\\
U_{ab} &=& -l_a \tilde m_b +l_b \tilde m_a \, .
\end{eqnarray}
where $m_a, \tilde m_b, k_a$ and $l_a$ are null tetrads. In the next
section we present an example of these kind of exact solutions.

%****************************************************
\section{Colliding wave solution in MAG}

Let us consider a MAG solution in the interaction region I, 
i.e. the region arising after the collision of the waves.
The coframe in the coordinates $(u,v,x,y)$ reads:
\begin{eqnarray}
\vartheta^{\hat{0}} & = & \sqrt{\Sigma} \, \left( \frac{du}{U} -
\frac{dv}{V}
 \right) \nonumber\\
\vartheta^{\hat{1}} & = & \sqrt{\Sigma} \, \left( \frac{du}{U} +
\frac{dv}{V}
\right) \nonumber\\
\vartheta^{\hat{2}} & = & \sqrt{\frac{\Delta}{\Sigma}} \, \left(dx +j^2\,
\left(uv-UV\right)^2 \, dy\right) \nonumber\\
\vartheta^{\hat{3}} &=& \sqrt{\frac{1}{\Sigma}}\,
\left(uv-UV\right)\,
 \left\{ j\, dx +\left(\left[\kappa m + {\widetilde a}
(uV-vU)\right]^2 +j^2 \right) dy \right\} \, ,
\label{frame2}
\end{eqnarray}
with two unknown functions $\Sigma(u,v), \Delta(u,v)$. Consequently,
the metric is given by
\begin{eqnarray}
g_{\hbox{\scriptsize (I)}} &=& 4\Sigma \, \frac{du}{U} \frac{dv}{V} +
\frac{1}{\Sigma}
\, \left\{ \left(uv-UV\right)^2 \left[ j dx + \left( j^2 +
\left[\kappa m +
{\widetilde a} (uV-vU) \right]^2 \right) dy \right]^2 \right.
 \nonumber \\
& & + \left. \Delta \left[ dx + j^2 (uv-UV)^2
dy\right]^2 \right\} \, ,
\end{eqnarray}
where
\begin{eqnarray}
U & := & \sqrt{1-u^2} \, , \qquad V:= \sqrt{1-v^2} \,  \nonumber\\
\Sigma & = & \left[ {\widetilde a}\left(uV-vU\right) + \kappa m
\right]^2
+ j^2 \left(uV+vU\right)^2\, , \nonumber\\
{\widetilde a}^2 &=& m^2 \kappa^2 - j^2 - q_1^2 \, , \nonumber\\
 \Delta &=& {\widetilde a}^2 (uv + UV)^2 \,
\end{eqnarray}
The nonmetricity and the torsion read as follows:
\begin{eqnarray}
Q^{\alpha\beta}_{\hbox{\scriptsize (I)}} & = &
\frac{{\widetilde a}(uV-vU)+\kappa m}{\sqrt{\Sigma \Delta}}
\,\left[k_0 N\, o^{\alpha\beta}
+\frac{4}{9}\,
k_1 N\,\left(\vartheta^{(\alpha}e^{\beta )}\rfloor-\frac{1}{4}\,
o^{\alpha\beta}\right)\right]\vartheta^{\hat{2}}\,,
\label{nichtmetrizitaetI} \\
T^\alpha_{\hbox{\scriptsize (I)}} & = & \frac{k_2 N}{3} \frac{{\widetilde
a}(uV-vU)+\kappa m}
{\sqrt{\Sigma \Delta}}  \,
\vartheta^\alpha\wedge\vartheta^{\hat{2}}
\label{tosionI}\,.
\end{eqnarray}
Here $j$, $m$, $q_1$ and $N$ are arbitrary {\em integration constants}, and 
the 
coefficients $k_{0}, k_{1}, k_{2}$ in the ansatz (\ref{genEug}) are determined
by the dimensionless coupling constants of the Lagrangian:
\begin{eqnarray}
k_0 &=& \left({a_2\over 2}-a_0\right)(8b_3 + a_0) - 3(c_3 + a_0)^2\,,
\label{k0}\\
k_1 &=& -9\left[ a_0\left({a_2\over 2} - a_0\right) +
(c_3 + a_0 )(c_4 + a_0 )\right]\,,
\label{k1}\\
k_2 &=& {3\over 2} \left[ 3a_0 (c_3 + a_0 ) + (
8b_3 + a_0)(c_4 + a_0 )\right]\,.\label{k2}
\end{eqnarray}
A rather weak condition, which must be imposed on these coefficients,
prescribes a value for the coupling constant $b_4$, namely
\begin{equation}
  b_4=\frac{a_0k+2c_4k_2}{8k_0}\,,\qquad\hbox{with}\qquad k:=
  3k_0-k_1+2k_2
\label{b4}\, .
\end{equation}
and the following relation for $z_4$
\begin{equation}
{q_1}^2=\kappa z_4 \frac{(k_0 N)^2}{2a_0}
\label{z4}\, .
\end{equation}

Our solution can be extended to the full spacetime by introducing 
the Heaviside step function 
\begin{equation}
\Theta (u)= \cases{1\, , \quad u \ge 0\cr
                  0\, , \quad u<0}\; ,
\end{equation} 
with $\Theta^2(u) = \Theta(u)$, 
and replacing $U \rightarrow \sqrt{1 - \Theta(u) u^2}$ and 
$V \rightarrow \sqrt{1 - \Theta(v) v^2}$, cf.\ \cite{taub}.

Then in region II the coframe reduces to
\begin{eqnarray}
\vartheta^{\hat{0}} & = & \sqrt{\Sigma}\, \left( du - \frac{dv}{V}
\right) \nonumber\\
\vartheta^{\hat{1}} & = & \sqrt{\Sigma} \, \left( du + \frac{dv}{V}
\right) \nonumber\\
\vartheta^{\hat{2}} & = & \sqrt{\frac{\Delta}{\Sigma}}\, (dx +j^2\,
\left(1-v^2\right)\, dy) \nonumber\\
\vartheta ^{\hat{3}}  &= & -\sqrt{\frac{1}{\Sigma}}\, \sqrt{ 1-v^2}
\left\{ j\, dx +\left[ \left(\kappa m -{\widetilde a} v\right)^2
+j^2\right] dy \right\}\, ,
\label{frame22}
\end{eqnarray}
and the corresponding metric is given by
\begin{eqnarray}
g_{\hbox{\scriptsize (II)}} &=& 4\Sigma \, \frac{du dv}{\sqrt{1-v^2}} +
\frac{1}{\Sigma}\, \left(1-v^2\right)\left\{ \left[j dx + [j^2 
+ \left(\kappa m -
{\widetilde a} v\right)^2] dy\right]^2 \right. \nonumber \\
& & + \left. {\widetilde a}^2 \left[ dx + j^2 (1-v^2)
dy\right]^2 \right\} \, ,
\end{eqnarray}
where
\begin{equation}
\Sigma = \left( \kappa m - {\widetilde a} v\right)^2 + j^2 v^2\, ,
\qquad  \Delta = {\widetilde a}^2 (1-v^2) \, ,
\end{equation}
which represents a plane wave solution (in the sense of Petrov
classification, a type N solution\cite{alberto,taub}).
The nonmetricity and the torsion in this region can be written as
follows:
\begin{eqnarray}
Q^{\alpha\beta}_{\hbox{\scriptsize (II)}} & = &
\frac{\kappa m - {\widetilde a} v}{\sqrt{\Sigma \Delta}}
\,\left[k_0 N\, o^{\alpha\beta}
+\frac{4}{9}\,
k_1 N\,\left(\vartheta^{(\alpha}e^{\beta )}\rfloor-\frac{1}{4}\,
o^{\alpha\beta}\right)\right]\vartheta^{\hat{2}}\,,
\label{nichtmetrizitaetII} \\ 
T^\alpha_{\hbox{\scriptsize (II)}} & = & \frac{k_2 N}{3} \frac{\kappa m - {\widetilde a} v} 
{\sqrt{\Sigma \Delta}}  \,
\vartheta^\alpha\wedge\vartheta^{\hat{2}}
\label{tosionII}\, .
\end{eqnarray}

In region III we  arrive at the
coframe
\begin{eqnarray}
\vartheta^{\hat{0}} &  & \sqrt{\Sigma}\, \left( \frac{du}{U} - dv
\right) \nonumber\\
\vartheta^{\hat{1}} & = & \sqrt{\Sigma} \, \left( \frac{du}{U} + dv
\right) \nonumber\\
\vartheta^{\hat{2}} & = & \sqrt{\frac{\Delta}{\Sigma}} \,(dx +j^2\,
\left(1-u^2\right) \, dy) \nonumber\\
\vartheta^{\hat{3}} & = & -\sqrt{\frac{1}{\Sigma}}\, \sqrt {1- u^2}
\left\{ j\, dx+ \left[\left(\kappa m + {\widetilde a} u\right)^2 +j^2\right] dy
\right\} \, ,
\label{frame23}
\end{eqnarray}
and the metric takes the following form:
\begin{eqnarray}
g_{\hbox{\scriptsize (III)}} & = & 4\Sigma \, \frac{du dv}{\sqrt{1-u^2}} +
\frac{1}{\Sigma}
\, \left(1-u^2\right)\left\{ \left[j dx + [j^2 + \left(\kappa m +
{\widetilde a} u\right)^2] dy\right]^2 \right. \nonumber \\
& & + \left. {\widetilde a}^2 \left[ dx + j^2 (1-u^2)
dy\right]^2 \right\} \, ,
\end{eqnarray}
where
\begin{equation}
\Sigma = \left( \kappa m + {\widetilde a} u\right)^2 + j^2 u^2\, ,
\qquad
 \Delta = {\widetilde a}^2 (1-u^2) \, .
\end{equation}
The nonmetricity and the torsion are now given by
\begin{eqnarray}
Q^{\alpha\beta}_{\hbox{\scriptsize (III)}} & = & 
\frac{\kappa m + {\widetilde a} u}{\sqrt{\Sigma \Delta}}
\,\left[k_0 N\, o^{\alpha\beta}
+\frac{4}{9}\,
k_1 N\,\left(\vartheta^{(\alpha}e^{\beta )}\rfloor-\frac{1}{4}\,
o^{\alpha\beta}\right)\right]\vartheta^{\hat{2}}\,,
\label{nichtmetrizitaetIII}
\\
T^\alpha_{\hbox{\scriptsize (III)}} & = & \frac{k_2 N}{3} \frac{\kappa m + {\widetilde a} u}
{\sqrt{\Sigma \Delta}}  \,
\vartheta^\alpha\wedge\vartheta^{\hat{2}}
\label{tosionIII}\, .
\end{eqnarray}
Here and in region II, $k_0$, $k_1$, and $k_2$ still satisfy
(\ref{k0}), (\ref{k1}) and (\ref{k2}).
It is easy to see that this is also a wave solution.

Finally, in the flat region IV
\begin{eqnarray}
\vartheta^{\hat{0}} & = & \sqrt{\Sigma}\, \left( du - dv \right) \nonumber\\
\vartheta^{\hat{1}} & = & \sqrt{\Sigma} \, \left( du + dv \right ) \nonumber\\
\vartheta^{\hat{2}} & = & \sqrt{\frac{\Delta}{\Sigma}}\, (dx +j^2\,
dy)\nonumber\\
\vartheta^{\hat{3}} & =& - \sqrt{\frac{1}{\Sigma}}\,  \left\{ j\, dx
+\left[ (km)^2+j^2\right] dy \right\} \label{frame24} \\
g_{\hbox{\scriptsize (IV)}} & = & 4\kappa^2 \, m^2 \, du \, dv + 
\frac{1}{(\kappa m)^2} \,
\left\{ \left[j dx + [j^2 + \left(\kappa m\right)^2] dy\right]^2
\right. + \left. {\widetilde a}^2 \left[ dx + j^2 dy\right]^2
\right\} \, ,
\end{eqnarray}
which is always reducible to the flat Minkowski form.

This solution was checked with Reduce \cite{REDUCE} with its Excalc
package \cite{EXCALC} for treating exterior differential forms\cite{Stauffer}
and the Reduce--based GRG computer algebra system \cite{GRG}.
The way of derivation of this solution is related to the search of a class 
of cylindrically symetric solutions in MAG, starting with the line element
\begin{equation}
ds^2= \Delta \left( \frac{dp^2}{P(p)} - \frac{dq^2}{Q(q)}\right) +
\frac{Q}{\Delta}\left(d\tau + \widetilde N(q)d\sigma\right)^2 
+ \frac{P}{\Delta}\left(d\tau + \widetilde M(p)d\sigma\right)^2\, , \nonumber
\end{equation}
with $\Delta:= \widetilde M - \widetilde N$. Assuming first that 
$P$ and $Q$ are polynomials up to fourth degree
on $p$ and $q$, respectively, second $\widetilde M$ and $\widetilde N$ are 
polynomials up to second degree also on $p$ and $q$, and third the 
torsion and nonmetricity are proportional
rational functions, then one arrives at algebraic equations, solvable by 
computer algebra programs, for the polynomials 
coefficients. It is always possible to introduce the $u$ and $v$ coordinates 
through $p=uV+vV$, $q=uV-vU$, $U=\sqrt{1-u^2}$, and $V=\sqrt{1-v^2}$.
However, only certain solutions satisfy the requirement of Ernst colliding 
waves (compare ref. 12 and section 3).

%**************************************************
\section{Discussion}

As it has been pointed out, the solution presented describes the
scattering of two noncollinear polarized gravitation plane waves.  At
the leading edge of each colliding type--N gravitational  wave, the
curvature tensor exhibits a jump discontinuity arising, for example, from the 
second derivative 
$(-U^2)^{\prime\prime} = u^2 \delta^\prime(u) + 4 u \delta(u) + 2 \Theta(u)$.  
The former is interpreted as a gravitational impulsive
wave, whereas the latter is attributed to a gravitational shock wave.

As far as the nonmetricity and torsion are concerned, if they are
considered as fundamental quantities then they behave as continuous
functions when crossing different regions; if they were considered as
secondary quantities defined by means of derivatives of more
fundamental functions, then they could present delta singularities and
jump discontinuities.
However, even then the Bianchi identities hold in a distributional sense, 
see \cite{taub}. 
In particular, also $D T^\alpha = R_\beta^{\phantom{\alpha}\alpha}\wedge 
\vartheta^\beta$ holds. 
There are no problems on the right--hand side because the delta type 
singularities of the curvature are multiplied by the smooth distributions 
$\sqrt{1 - \Theta(u) u^2}$ and $\sqrt{1 - \Theta(v) v^2}$, respectively.

So far it is not quite clear, if this special MAG model has problems with 
{\em redundant variables}.
In the case of restricted Poincar\'e gauge models (without nonmetricity),  
a similar reduction (induced via a double duality ansatz)
was based on the {\em teleparallelism equivalence}, see Baekler et 
al. Ref.  \cite{bm86}. However, it was shown by Lenzen \cite{{le84}}, and 
later confirmed in Ref. \cite{bm88} that then necessarily {\em free} functions
occur in exact torsion solutions. (The tentavive gauge fixing approach 
suggested there as a way out met considerable criticism.) 
Thus for the so--called ``viable" set there exist 
{\em infinite} many exact vacuum solutions 
which may indicate a physically problematic {\em degeneracy} of those 
models. 
Recent reports to rescue the initial value problem in PG theory
by Hecht et al. \cite{he96} and the Refs. therein, seem not to be 
conclusive. 

The related situation for MAG is not yet resolved, since again 
a {\em teleparallelism type relation}, see (5.9.16) of Ref. \cite{PR},
seems to be crucial for the equivalence proof of MAG with the Einstein--Proca 
Lagrangian. 
Already earlier, within the framework of the Poincar\'e gauge theory
(PG) of gravity, the post--Newtonian generation of gravitational radiation
in one parameter teleparallelism type $T^2$ models were studied by Schweizer 
et al. \cite{Schw}. In a first order approximation no deviation from Einstein's
GR was found; also, as a bonus, the dipole gravitational radiation of other 
alternative theories is absent here.

More recently, plane wave solutions of GR are generalized to $R+R^2
+T^2$ models by Zhytnikov \cite{zhyt}. (Note here the possibility of
notational confusion, since $Q$ is there used for torsion.) Our present paper 
is an extension of this work in two different directions: First we extend
to models with nonmetricity $Q_{\alpha\beta}$ including the Weyl
covector $Q$ and, secondly, also colliding waves exhibiting shock
fronts are considered.

\acknowledgments

We would like to thank Friedrich W. Hehl for useful discussions and literature
hints.
This work was partially supported by  CONACyT, grants No.
3544--E9311, No. 3898P--E9608, No. 3692P--E9607 and by the joint 
German--Mexican project KFA--Conacyt
E130--2924 and DLR--Conacyt 6.B0a.6A. Moreover, E.W.M. acknowledges
the support by the short--term fellowship 961 616 015 6 of the German
Academic Exchange Service (DAAD), Bonn, and C.L. thanks the Deutsche
Forschungsgemeinschaft and the DAAD for financial support.

\newpage

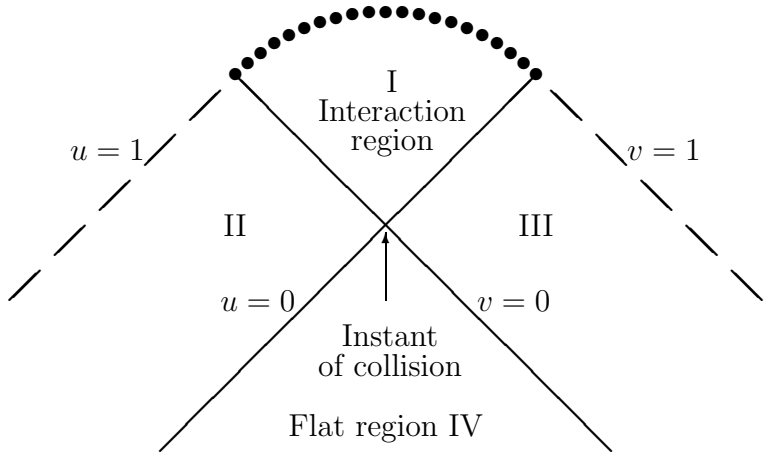
\begin{figure}
\unitlength1cm
\begin{picture}(10,7)
\put(3.3,2){\makebox(0,0){$u = 0$}}
\put(6.7,2){\makebox(0,0){$v = 0$}}
\put(1.3,4){\makebox(0,0){$u = 1$}}
\put(8.7,4){\makebox(0,0){$v = 1$}}
\put(5,2){\vector(0,1){0.9}}
\put(4.1,1){\shortstack[c]{Instant \\ of collision}}
\put(5,0.3){\makebox(0,0){Flat region IV}}
\put(4.1,4){\shortstack[c]{I \\ Interaction \\ region}}
\put(3,3){\makebox(0,0){II}}
\put(7,3){\makebox(0,0){III}}
\put(7,5){\makebox(0,0){$\bullet$}}
\put(6.837,5.151){\makebox(0,0){$\bullet$}}
\put(6.663,5.288){\makebox(0,0){$\bullet$}}
\put(6.478,5.412){\makebox(0,0){$\bullet$}}
\put(6.284,5.52){\makebox(0,0){$\bullet$}}
\put(6.082,5.613){\makebox(0,0){$\bullet$}}
\put(5.874,5.69){\makebox(0,0){$\bullet$}}
\put(5.66,5.75){\makebox(0,0){$\bullet$}}
\put(5.443,5.794){\makebox(0,0){$\bullet$}}
\put(5.222,5.82){\makebox(0,0){$\bullet$}}
\put(5,5.828){\makebox(0,0){$\bullet$}}
\put(4.778,5.82){\makebox(0,0){$\bullet$}}
\put(4.558,5.794){\makebox(0,0){$\bullet$}}
\put(4.34,5.75){\makebox(0,0){$\bullet$}}
\put(4.126,5.69){\makebox(0,0){$\bullet$}}
\put(3.918,5.613){\makebox(0,0){$\bullet$}}
\put(3.716,5.52){\makebox(0,0){$\bullet$}}
\put(3.522,5.412){\makebox(0,0){$\bullet$}}
\put(3.337,5.288){\makebox(0,0){$\bullet$}}
\put(3.163,5.151){\makebox(0,0){$\bullet$}}
\put(3,5){\makebox(0,0){$\bullet$}}
\thicklines
\multiput(0,2)(0.6,0.6){5}{\line(1,1){0.4}}
\put(2,0){\line(1,1){5}}
\multiput(10,2,0)(-0.6,0.6){5}{\line(-1,1){0.4}}
\put(8,0){\line(-1,1){5}}
\end{picture}

\bigskip
\bigskip
\bigskip
\caption{
The four regions of the spacetime:
Region IV where the waves propagate is flat. The impulsive
gravitational
waves propagate along the null boundaries $v=0$ and $u=0$, separating
regions II and IV, and III and IV, respectively. In region II,
observers see the shower of pure gravitational radiation following the wave 
front propagating along $v=0$. Symmetrical consideration apply in region III.
The collision occurs at (0,0) and the interaction is described by region I.
 }
\end{figure}


\begin{references}

\bibitem{khan} K. Khan and R. Penrose, {\em Nature} {\bf 229} (1971) 185.

\bibitem{szek} P. Szekeres, {\em J. Math. Phys.} {\bf 13} (1972) 286.

\bibitem{chandra1} S. Chandrasekhar and V. Ferrari, {\em Proc. R. Soc. London}
Ser. {\bf A396} (1984) 55.

\bibitem{chandra2} S. Chandrasekhar and B.C. Xanthopoulos, {\em Proc.
R. Soc. London} Ser. {\bf A398} (1985) 223.

\bibitem{PR} F.W. Hehl, J.D. McCrea, E.W.  Mielke, and Y. Ne'eman,
{\em Phys. Rep.} {\bf 258} (1995) 1.

\bibitem{Tres14} R. Tresguerres, {\em Z. Phys.} {\bf C65} (1995) 347.

\bibitem{Tres15} R. Tresguerres, {\em Phys. Lett.} {\bf A200} (1995) 405.

\bibitem{TW} R.W.  Tucker and C.  Wang, {\em Class. Quantum Grav.} {\bf 12} 
(1995) 2587.

\bibitem{De96}
T. Dereli, M. \"Onder, J. Schray, R.W. Tucker, and C. Wang,
{\em Class. Quantum Grav.} {\bf 13} (1996) L103.

\bibitem{ma} A. Mac\'{\i}as, E.W. Mielke and J. Socorro, {\em Class. Quantum
Grav.} (1997). In print.

\bibitem{slmm} J. Socorro, C. L\"ammerzahl, A. Mac\'{\i}as, and E.W. Mielke:
``Multipole solutions in metric--affine gravity". (1997) Submitted.

\bibitem{ernst} F.J. Ernst, A. Garc\'{\i}a, and I. Hauser,
{\em J. Math. Phys.} {\bf 28} (1987) 2951.

\bibitem{alberto1} A. Garc\'{\i}a and N. Bret\'on, {\em Phys. Rev. } {\bf D53}
(1996) 4351.

\bibitem{alberto} A. Garc\'{\i}a, {\em J. Math. Phys.} {\bf 25}
(1984) 1951.

\bibitem{taub} A.H. Taub, {\em J. Math. Phys.} {\bf 21} (1980) 1423.

\bibitem{REDUCE} A.C. Hearn, {\em REDUCE User's Manual. Version 3.6}.
Rand publication CP78 (Rev.\ 7/95) (RAND, Santa Monica, CA 90407-2138, USA, 
1995).

\bibitem{EXCALC} E. Schr\"ufer, F.W. Hehl, and J.D. McCrea, {\em Gen. Relat. 
Grav.} {\bf 19} (1987) 197.

\bibitem{Stauffer} D. Stauffer, F.W. Hehl, N. Ito, V. Winkelmann, and
J.G. Zabolitzky: {\em Computer Simulation and Computer Algebra -- Lectures for
Beginners.} 3rd ed.\  (Springer, Berlin, 1993).

\bibitem{GRG} V.V. Zhytnikov, {\em GRG. Computer Algebra System for
Differential Geometry, Gravity and Field Theory. Version 3.1} (Moscow, 1991) 
108 pages.

\bibitem{Schw}M. Schweitzer, N. Straumann, and A. Wipf, {\em Gen.
Rel. Grav.} {\bf 12} (1980) 951.

\bibitem{he96} R.D. Hecht, J.M. Nester, and V.V. Zhytnikov, {\em Phys. Lett.} 
{\bf A222} (1996) 37.

\bibitem{le84} J. Lenzen, {\em Nuovo Cim.} {\bf B82} (1984) 85.

\bibitem{bm86} P. Baekler and E.W. Mielke, {\em Phys. Letters} {\bf 113A} 
(1986) 471.

\bibitem{bm88} P. Baekler and E.W. Mielke, {\em Fortschr. Phys.} {\bf 36}
(1988) 549.

\bibitem{Kaku}M. Kaku: {\em Quantum Field Theory} (Oxford University
Press 1993). p.  750.

\bibitem{zhyt}V.V. Zhytnikov, {\em J. Math. Phys.} {\bf 35} (1994) 6001.

\end{references}
\end{document}